\documentclass[preprint,showpacs,amsmath,amssymb,eqsecnum]{revtex4}
\usepackage{dcolumn}
\usepackage{bm}
\usepackage{graphicx}

\begin{document}
\title{{Quark Mixing from Mass Matrix Model with Flavor 2 \(\leftrightarrow\) 3 Symmetry}}

\author{Hiroyuki NISHIURA}
\affiliation{%
Faculty of Information Science and Technology, 
Osaka Institute of Technology, 
Hirakata, Osaka 573-0196, Japan}

\author{Koichi MATSUDA}
\affiliation{%
Center for High Energy Physics, 
Department of Engineering Physics,  
Tsinghua University, Beijing 100084, China
\footnote{Present address: 1-13 Akasaka, Kinugasa, Kyoto 603-8486, Japan} 
}

\author{Takeshi FUKUYAMA}
\affiliation{%
Department of Physics and Global Innovation Research Organization, Ritsumeikan University,
Kusatsu, Shiga, 525-8577, Japan}

\date{September 25, 2008}

\begin{abstract}
We consider an universal mass matrix model which has a seesaw-invariant structure 
with the most general texture based on flavor 2 \(\leftrightarrow\) 3 symmetry 
common to all quarks and leptons.  
The CKM quark mixing matrix of the model is analyzed.
It is shown that the model is consistent with all the experimental data of quark mixings by tuning free parameters of the model.  
We also show that the values of parameters of the present model consistent with the experimental data 
are not far from the ones of the mass matrix model with a vanishing (1,1) element.

\end{abstract}


\maketitle
\section{Introduction}
The discovery of neutrino oscillation~\cite{skamioka}--\cite{kamland} indicates that 
neutrinos have finite masses and mix one another with near tri-bimaximal lepton mixings 
in contrast to small quark mixings.  
In order to explain the large lepton mixings and small quark mixings, 
mass matrix models with various structures such as zero texture~\cite{fritzsch}--\cite{Plentinger}, 
flavor $2 \leftrightarrow 3$ symmetry~\cite{Fukuyama}--\cite{Matsuda5} etc. 
have been investigated in the literature. 
Based on an idea that quarks and leptons should be unified, 
it is an interesting approach to investigate a possibility that 
all the mass matrices of the quarks and leptons have the same form  
which can lead to the large lepton mixings and the small quark mixings simultaneously.  
Since the mass matrix model is intended to be embedded into a grand unified theory (GUT),  
it is also desirable for the model to have the following features:  
(i) The structure is common to all the mass matrices,  
\(M_u\), \(M_d\), \(M_e\), and \(M_\nu \) for up quarks (\(u,c,t\)), down quarks (\(d,s,b\)),  
charged leptons (\(e,\mu,\tau \)), and  neutrinos (\(\nu_e,\nu_{\mu},\nu_{\tau} \)), respectively.  
(ii) Since we assume the seesaw mechanism~\cite{Minkowski}--\cite{Schechter2} for neutrino masses,
the structure should conserve its form through the relation \(M_\nu \simeq M_L-M_D M_R^{-1} M_D^T\).
We call this structure as a seesaw-invariant form~\cite{Fukuyama2}. Here \(M_D\),\(M_L\) and \(M_R\) are, respectively, the Dirac, the left- and the right-handed Majorana type neutrino mass matrices, 
which are also assumed to have the same structure.
\par
In our previous works \cite{Koide,Matsuda,Matsuda2,Matsuda3,Matsuda4,Matsuda5}, we investigated a mass matrix model 
based on the flavor $2 \leftrightarrow 3$ symmetry. 
We pointed out that this approach leads to reasonable values for the small quark mixing 
as well as the large lepton mixing, and that  
the same texture form can give a universal description of quark and lepton mass matrices.
In those works, we assume that (1,1) element of the mass matrix is zero. 
However, it is not clear that vanishing (1,1) element is necessary. 
From a phenomenological point of view, it is preferable~\cite{Fukuyama3} for mass matrix to have 
as many components as possible based on flavor $2 \leftrightarrow 3$ symmetry 
in order to be embedded into a GUT.  
If the experimental data prefer a vanishing (1,1) element, there must be some discrete symmetry~\cite{Koide}.  
\par
In this paper, as typical mass matrices which have the features mentioned above, 
we consider quark mass matrices $M_f$ for $f=u$ and $d$ with a nonvanishing (1,1) element 
and  most general $2 \leftrightarrow 3$ symetric form given by 
\begin{equation}
M_f=P^\dagger_f \widehat{M_f} P_f \quad \mbox{for } f=u \mbox{ and } d.\\ 
\end{equation}
Here $P_f$ is a diagonal phase matrix given by 
\begin{equation}
P_f =\mbox{diag}\left(e^{i\alpha_{f1}}, e^{i\alpha_{f2}}, e^{i\alpha_{f3}}\right).
\end{equation}
Here we consider a nonsymmetric matrix $\widehat{M_f}$ which is the most general form 
based on the $2 \leftrightarrow 3$ symmetry  defined by
\begin{equation}
\widehat{M_f}=\left(
\begin{array}{lll}
\ D_f & \ A_f & \ A_f \\
\ A^\prime_f & \ B_f & \ C_f \\
\ A^\prime_f & \ C_f & \ B_f \\
\end{array}
\right), \ \label{M_hat_f}
 \end{equation}
where $A_f$, $A^\prime_f$, $B_f$, $C_f$, and $D_f$ are real parameters. 
It is noted that the diagonal phase matrix $P_f$ which breaks $2 \leftrightarrow 3$ symmetry 
has been introduced from a phenomenological point of view.
We believe that $2 \leftrightarrow 3$ symmetry or more higher $A_4$ is essentially concerned 
with real matrix. 
These have nice characters as a dominant part at least in the lepton sector.
However, CP violating phases are very important in the lepton sector as well as in the quark sector.
We present analytical expressions for the Cabibbo-Kobayashi-Maskawa (CKM) 
quark mixing matrix~\cite{CKM1}--\cite{CKM2} of the model and investigate 
whether the model is consistent or not with the experimental data.
\par
This article is organized as follows. 
In Sec.~II, we discuss the diagonalization of mass matrix of our model. 
In Sec.~III, analytical expressions of the quark mixing matrix of the model are given. 
Data fitting of the CKM quark mixing matrix of the model is also discussed.
Sec.~IV  is devoted to a conclusion and discussion.

\section{Diagonalization of Mass matrix}
\par
We discuss a diagonalization of the mass matrix $M_f$. 
First we argue a diagonalization of $\widehat{M_f}$ given by Eq.~(\ref{M_hat_f}). 
This is diagonalized  by two orthogonal matrices $O_{f1}$ and $O_{f2}$ as 
\begin{equation}
O_{f1}^T \widehat{M_f} O_{f2} = \mbox{diag}(m_{f1} , m_{f2} , m_{f3}),
\end{equation}
where $m_{f1}$, $m_{f2}$, and $m_{f3}$ are eigenvalues of $M_f$. The orthogonal matrices $O_{f1}$ and $O_{f2}$ are given by 
\begin{equation}
O_{f1} = U \left(
\begin{array}{ccc}
\cos\varphi_{f1} & -\sin\varphi_{f1} &  0 \\
\sin\varphi_{f1} & \cos\varphi_{f1} &  0 \\
0 & 0 & 1
\end{array}
\right),\label{O_f_1}
\end{equation}
\begin{equation}
O_{f2} = U \left(
\begin{array}{ccc}
\cos\varphi_{f2} & -\sin\varphi_{f2} &  0 \\
\sin\varphi_{f2} & \cos\varphi_{f2} &  0 \\
0 & 0 & 1
\end{array}
\right),\label{O_f_2}
\end{equation}
where $U$ is a tri-bimaximal mixing matrix given by
\begin{equation}
U = \left(
\begin{array}{ccc}
 \frac{2}{\sqrt{6}} &  \frac{1}{\sqrt{3}} &  0 \\
- \frac{1}{\sqrt{6}}  & \frac{1}{\sqrt{3}}  &  - \frac{1}{\sqrt{2}}  \\
- \frac{1}{\sqrt{6}} & \frac{1}{\sqrt{3}}  & \frac{1}{\sqrt{2}} 
\end{array}
\right).\label{tri-bimixing_matrix}
\end{equation}
Here we parameterize $\varphi_{f1}$ and $\varphi_{f2}$ as sifts of $O_{f1}$ and $O_{f2}$ from a tri-bimaximal mixing matrix.
\par 
We have five component parameters in $\widehat{M_f}$, namely, $A_f$, $A^\prime_f$, $B_f$, $C_f$, and $D_f$.
If we fix the $m_{fi}$ by the observed quark mass, 
we have two free parameter left. Therefore we choose $\varphi_{f1}$ and $\varphi_{f2}$ as the free parameters. 
Namely, the components $A_f$, $A^\prime_f$, $B_f$, $C_f$, and $D_f$ in Eq.~(\ref{M_hat_f}) are presented 
in terms of $\varphi_{f1}$ and $\varphi_{f2}$ as follows:
\begin{eqnarray}
D_f &=& m_{f1} \cos(\varphi_{f1}-\alpha) \cos(\varphi_{f2}-\alpha) + m_{f2}\sin(\varphi_{f1}-\alpha) \sin(\varphi_{f2}-\alpha)\ ,\\
A_f &=& \frac{1}{\sqrt{2}} \left[ m_{f1} \cos(\varphi_{f1}-\alpha) \sin(\varphi_{f2}-\alpha) - m_{f2}\sin(\varphi_{f1}-\alpha) \cos(\varphi_{f2}-\alpha) \right]\ ,\\
A^\prime_f &=& \frac{1}{\sqrt{2}} \left[ m_{f1} \sin(\varphi_{f1}-\alpha) \cos(\varphi_{f2}-\alpha) - m_{f2}\cos(\varphi_{f1}-\alpha) \sin(\varphi_{f2}-\alpha) \right]\ ,\\
B_f &=& \frac{1}{2} \left[ m_{f1} \sin(\varphi_{f1}-\alpha) \sin(\varphi_{f2}-\alpha) + m_{f2}\cos(\varphi_{f1}-\alpha) \cos(\varphi_{f2}-\alpha) + m_{f3} \right]\ ,\\
C_f &=& \frac{1}{2} \left[ m_{f1} \sin(\varphi_{f1}-\alpha) \sin(\varphi_{f2}-\alpha) + m_{f2}\cos(\varphi_{f1}-\alpha) \cos(\varphi_{f2}-\alpha) - m_{f3} \right]. 
\end{eqnarray}
Here $\alpha $ is a constant angle defined by 
\begin{equation}
\tan\alpha \equiv \frac{1}{\sqrt{2}}.
\end{equation}
\section{CKM quark mixing matrix}
Let us discuss the quark mixing matrix.
The mass matrices $M_u$ and $M_d$ for the u- and d-quarks are, respectively, given by
\begin{eqnarray}
M_u&=&P^\dagger_u \widehat{M_u}P_u, \\
M_d&=&P^\dagger_d \widehat{M_d}P_d,
\end{eqnarray}
where $P_{u}$ and $P_{d}$ are diagonal phase matrices and  
$\widehat{M_u}$ and $\widehat{M_d}$ are given by Eq.~(\ref{M_hat_f}). 
The mass matrix $M_f$ ($f=u\mbox{ and }d$) are diagonalized as 
\begin{equation}
U_{Lf}^\dagger M_f U_{Rf} = \mbox{diag}\left(m_{f1}, m_{f2}, m_{f3}\right).
\end{equation}
The unitary matrix $U_{Lf}$ and $U_{Rf}$ are, respectively, described as
\begin{eqnarray}
U_{Lf} &=&P_{f}^\dagger O_{f1},\\
U_{Rf} &=&P_{f}^\dagger O_{f2}.
\end{eqnarray}
Therefore the CKM quark mixing matrix $U_{CKM}$ 
of the model is given by 
\begin{equation}
U_{CKM}=U^\dagger_{Lu}U_{Ld}=O^{T}_{u1} PO_{d1},\label{our_ckm}
\end{equation}
where $P \equiv P_uP^\dagger_d$ is diagonal phase matrix given by
\begin{equation}
P 
=\mbox{diag}(e^{i(\alpha_{u1}-\alpha_{d1})} , e^{i(\alpha_{u2}-\alpha_{d2})} , e^{i(\alpha_{u3}-\alpha_{d3})})
\equiv \mbox{diag}(e^{i\tau} , e^{i\sigma}, 1).
\end{equation}
Here we take $\alpha_{d3}=\alpha_{u3}=0$ without any loss of generality.
\par
By using the expressions of $O_{d1}$ and $O_{u1}$ in Eq.~(\ref{O_f_1}), the explicit $(i,j)$ elements of $U_{CKM}$ 
are obtained as
\begin{eqnarray}
(U_{CKM})_{12} &=& 
-\left(\frac{2}{3}e^{i \tau}+\frac{1}{6}e^{i \sigma}+\frac{1}{6}\right) \cos\varphi_{u1} \sin\varphi_{d1} 
+\left(\frac{1}{3}e^{i \tau}+\frac{1}{3}e^{i \sigma}+\frac{1}{3}\right)\sin\varphi_{u1}\cos\varphi_{d1}\nonumber\\
&&+\left(\frac{\sqrt{2}}{3}e^{i \tau}-\frac{\sqrt{2}}{6}e^{i \sigma}+\frac{\sqrt{2}}{6}\right)\cos(\varphi_{u1}+\varphi_{d1}), \label{our_ckm_12}\\
(U_{CKM})_{22} &=& 
\left(\frac{2}{3}e^{i \tau}+\frac{1}{6}e^{i \sigma}+\frac{1}{6}\right) \sin\varphi_{u1} \sin\varphi_{d1} 
+\left(\frac{1}{3}e^{i \tau}+\frac{1}{3}e^{i \sigma}+\frac{1}{3}\right)\cos\varphi_{u1}\cos\varphi_{d1}\nonumber\\
&&-\left(\frac{\sqrt{2}}{3}e^{i \tau}-\frac{\sqrt{2}}{6}e^{i \sigma}+\frac{\sqrt{2}}{6}\right)\sin(\varphi_{u1}+\varphi_{d1}),  \label{our_ckm_22}\\
(U_{CKM})_{13} &=& \frac{\sqrt{3}}{6}  \left(e^{i \sigma}-1\right) \left[
\cos\varphi_{u1}- \sqrt{2} \sin\varphi_{u1} \right] \nonumber\\
&=&\frac{1}{\sqrt{3}} e^{i(\frac{\pi}{2}+\frac{\sigma}{2})} \sin\frac{\sigma}{2} \left[\cos\varphi_{u1}  - \sqrt{2}\sin\varphi_{u1} \right], \label{our_ckm_13}\\
(U_{CKM})_{23} &=& - \frac{\sqrt{3}}{6}  \left(e^{i \sigma}-1\right) \left[
 \sin\varphi_{u1}+ \sqrt{2}  \cos\varphi_{u1} \right] \nonumber\\
&=&-\frac{1}{\sqrt{3}} e^{i(\frac{\pi}{2}+\frac{\sigma}{2})} \sin\frac{\sigma}{2} \left[\sin\varphi_{u1} + \sqrt{2}\cos\varphi_{u1}  \right],  \label{our_ckm_23}\\
(U_{CKM})_{33} &=& \frac{1}{2}(e^{i \sigma}+1)=e^{i\frac{\sigma}{2}}\cos\frac{\sigma}{2},  \label{our_ckm_33}\\
(U_{CKM})_{31} &=&
\frac{\sqrt{3}}{6} \left(e^{i \sigma} -1\right) \left[\cos\varphi_{d1}- \sqrt{2} \sin\varphi_{d1}\right]\nonumber\\
&=&\frac{1}{\sqrt{3}} e^{i(\frac{\pi}{2}+\frac{\sigma}{2})}\sin\frac{\sigma}{2} \left[\cos\varphi_{d1}- \sqrt{2} \sin\varphi_{d1}\right],  \label{our_ckm_31}\\
(U_{CKM})_{32} &=&
-\frac{\sqrt{3}}{6} \left(e^{i \sigma} -1\right) \left[\sin\varphi_{d1}+ \sqrt{2} \cos\varphi_{d1}\right]\nonumber\\
&=&-\frac{1}{\sqrt{3}} e^{i(\frac{\pi}{2}+\frac{\sigma}{2})}\sin\frac{\sigma}{2} \left[\sin\varphi_{d1}+ \sqrt{2} \cos\varphi_{d1}\right],  \label{our_ckm_32}
\end{eqnarray}
\par
It should be noted that the above expressions of $(U_{CKM})_{ij}$ do not depend on the quark masses $m_{ui}$ and $m_{di}\ (i=1,2,3)$ of up and down quarks, respectively,
which we denoted  as $(m_u, m_c, m_t)$ and $(m_d, m_s, m_b)$ and fix by the observed masses.
Namely, only two component parameters $\varphi_{u1}$ and $\varphi_{d1}$ and two phase parameters $\tau$ and $\sigma$ are left as free parameters in the above expressions of $(U_{CKM})_{ij}$. 
Using this feature of the model, we can reproduce the observed data for $(U_{CKM})_{ij}$  as will be shown later. 
\par
By using the rephasing of the up and down quarks, 
Eq.~(\ref{our_ckm}) is changed to the standard representation of the CKM quark mixing matrix, 
\begin{eqnarray}
U_{CKM}^{\rm std} &=& \mbox{diag}(e^{i\zeta_1^u},e^{i\zeta_2^u},e^{i\zeta_2^u})  \ U_{CKM} \ 
\mbox{diag}(e^{i\zeta_1^d},e^{i\zeta_2^d},e^{i\zeta_2^d}) \nonumber \\
&=&
\left(
\begin{array}{ccc}
c_{13}^qc_{12}^q & c_{13}^qs_{12}^q & s_{13}^qe^{-i\delta_q} \\
-c_{23}^qs_{12}^q-s_{23}^qc_{12}^qs_{13}^q e^{i\delta_q}
&c_{23}^qc_{12}^q-s_{23}^qs_{12}^qs_{13}^q e^{i\delta_q} 
&s_{23}^qc_{13}^q \\
s_{23}^qs_{12}^q-c_{23}^qc_{12}^qs_{13}^q e^{i\delta_q}
 & -s_{23}^qc_{12}^q-c_{23}^qs_{12}^qs_{13}^q e^{i\delta_q} 
& c_{23}^qc_{13}^q \\
\end{array}
\right) \ .
\label{stdrep}
\end{eqnarray}
Here \(\zeta_i^q\) comes from the rephasing in the quark fields 
to make the choice of phase convention.
By using the expressions of $U_{CKM}$ in Eqs.~(\ref{our_ckm_12})-(\ref{our_ckm_23}), 
the $CP$ violating phase $\delta_q$ in the quark mixing matrix is given by
\begin{equation}
\delta_q =
\mbox{arg}\left[
\left(\frac{(U_{CKM})_{12} (U_{CKM})_{22}^{*}}{(U_{CKM})_{13} (U_{CKM})_{23}^{*}}\right) + 
\frac{|(U_{CKM})_{12}|^2}{1-|(U_{CKM})_{13}|^2}
\right]. \label{our_delta_q} \\
\end{equation}
\par
Thus we have obtained the analytical expressions for $|(U_{CKM})_{12}|$, $|(U_{CKM})_{23}|$, $|(U_{CKM})_{13}|$, and $\delta_q$ of the model 
which are given by Eqs.~(\ref{our_ckm_12}), (\ref{our_ckm_23}), (\ref{our_ckm_13}), and (\ref{our_delta_q}), 
respectively, as functions of the four parameters $\tau$, $\sigma$, $\varphi_{u1}$, and $\varphi_{d1}$. 
\par
On the other hand, the numerical values of $|(U_{CKM})_{12}|$, $|(U_{CKM})_{23}|$, $|(U_{CKM})_{13}|$, and $\delta_q$ 
at the unification scale \(\mu=M_X\)  are estimated from the experimental data observed at electroweak scale \(\mu=M_Z\) 
by using the renormalization group equation as~\cite{Matsuda3,Matsuda4}:
\begin{eqnarray}
|(U_{CKM})_{12}|&=&0.2226-0.2259,\label{U_CKM_BD_12}\\
|(U_{CKM})_{23}|&=&0.0295-0.0387,\label{U_CKM_BD_23}\\
|(U_{CKM})_{13}|&=&0.0024-0.0038,\label{U_CKM_BD_13}\\
\delta_q &=& 46^\circ   - 74^\circ ,\label{U_CKM_BD_delta} \\ 
|(U_{CKM})| &=& 
\left(
\begin{array}{ccc}
0.9741-0.9749 & 0.2226-0.2259 & 0.0024-0.0038 \\
0.2225-0.2259 & 0.9734-0.9745 & 0.0295-0.0387 \\
0.0048-0.0084 & 0.0289-0.0379 & 0.9993-0.9996
\end{array}
\right). \label{eq1118-01} 
\end{eqnarray}
The ratios among CKM matrix elements are 
\begin{align}
|(U_{CKM})_{21}/(U_{CKM})_{22}| &= 0.2284-0.2320, & 
|(U_{CKM})_{12}/(U_{CKM})_{22}| &= 0.2285-0.2321, \nonumber \\
|(U_{CKM})_{31}/(U_{CKM})_{32}| &= 0.1699-0.2252, & 
|(U_{CKM})_{31}/(U_{CKM})_{33}| &= 0.0050-0.0084, \nonumber \\
|(U_{CKM})_{13}/(U_{CKM})_{23}| &= 0.0747-0.1055, & 
|(U_{CKM})_{13}/(U_{CKM})_{33}| &= 0.0024-0.0038. \label{vratio}
\end{align}
In the numerical calculations, we use the running quark masses which is estimated 
in Ref.~\cite{Fusaoka}
[minimal supersymmetric standard model with tan$\beta$=10 case] : 
\begin{equation}
\begin{array}{lll}
m_u(m_Z)=2.33^{+0.42}_{-0.45}\, \mbox{MeV},& 
m_c(m_Z)=677^{+56}_{-61}\, \mbox{MeV},&
m_t(m_Z)=181\pm13\, \mbox{GeV},\\
m_d(m_Z)=4.69^{+0.60}_{-0.66}\, \mbox{MeV},& 
m_s(m_Z)=93.4^{+11.8}_{-13.0}\, \mbox{MeV},&
m_b(m_Z)=3.00\pm0.11\, \mbox{GeV},\\ 
\end{array}
\label{eq123103}
\end{equation}
\begin{equation}
\begin{array}{lll}
m_u(M_X)=1.04^{+0.19}_{-0.20}\, \mbox{MeV},& 
m_c(M_X)=302^{+25}_{-27}\, \mbox{MeV},&
m_t(M_X)=129^{+196}_{-40}\,  \mbox{GeV},\\
m_d(M_X)=1.33^{+0.17}_{-0.19}\, \mbox{MeV},& 
m_s(M_X)=26.5^{+3.3}_{-3.7}\, \mbox{MeV},&
m_b(M_X)=1.00\pm0.04\, \mbox{GeV}.\\ 
\end{array}
\label{eq123104}
\end{equation}
\par
By using the above experimental constraints as inputs, 
we obtain consistent solutions for the parameter $\tau$, $\sigma$, $\varphi_{u1}$, and $\varphi_{d1}$ of our model 
from our exact CKM matrix elements given by Eqs.~(\ref{our_ckm_12}), (\ref{our_ckm_23}), (\ref{our_ckm_13}), and (\ref{our_delta_q}). 
\par
From the expressions of $|(U_{CKM})_{31}|$ and $|(U_{CKM})_{32}|$ 
in Eqs.~(\ref{our_ckm_31}) and (\ref{our_ckm_32}),
we obtain the following constraint for the parameter $\varphi_{d1}$, 
which holds irrespectively of the free phase parameters $\tau$ and $\sigma$ and also $\varphi_{u1}$.
\begin{equation}
\frac{|(U_{CKM})_{31}|}{|(U_{CKM})_{32}|}=\left| \frac{ \cos\varphi_{d1}-\sqrt{2}\sin\varphi_{d1} }{\sin\varphi_{d1}+\sqrt{2}\cos\varphi_{d1}}\right| .
\end{equation}
\par
In doing parameter fitting, we first derive allowed region in the plane of the parameters $\sigma$ and $\varphi_{d1}$ from
the expressions of $|(U_{CKM})_{31}|$, $|(U_{CKM})_{32}|$ and $|(U_{CKM})_{33}|$ 
in Eqs.~(\ref{our_ckm_31}), (\ref{our_ckm_32}), and (\ref{our_ckm_33}) with using the experimental constraints given in Eqs.~(\ref{eq1118-01}). 
We obtain the allowed region in the $\sigma$ - $\varphi_{d1}$ plane shown in Fig.~1, 
from which it turns out that the following values for the parameters $\sigma$ and $\varphi_{d1}$ 
are consistent with the experimental data: 
\begin{eqnarray}
\sigma &\simeq&0.06-0.07 , \label{sigma_value}\\
\varphi_{d1}  &\simeq& 0.40-0.44\quad  \mbox{or} \quad 0.79-0.84.\label{var_phi_value}
\end{eqnarray}
\par
Next we obtain allowed region in the plane of remaining two parameters $\tau$ and $\varphi_{u1}$ from 
the constraints of $|(U_{CKM})_{12}|$, $|(U_{CKM})_{23}|$, $|(U_{CKM})_{13}|$, and $\delta_q$ 
given in Eqs.~(\ref{our_ckm_12}), (\ref{our_ckm_23}), (\ref{our_ckm_13}), and (\ref{our_delta_q}) 
with using the experimental constraints given in Eqs.~(\ref{U_CKM_BD_12})-~(\ref{U_CKM_BD_delta}).  
We obtain the allowed region shown in Fig.~2 taking the following values for $\sigma$ and $\varphi_{d1}$, 
$\sigma \simeq0.07$ and $\varphi_{d1}\simeq0.42$ or $0.81$. 
As seen from Fig.~2, it is found that the model is consistent with the experimental data for 
\begin{eqnarray}
\tau &\simeq& -(1.90 -1.55) \quad \mbox{or} \quad  1.25-1.57 ,\\ 
\varphi_{u1} &\simeq& 0.507-0.547 \quad \mbox{or} \quad 0.685-0.725.
\end{eqnarray}

\section{conclusion and discussion}
We consider the following mass matrix model for quarks with an universal form given by
$M_f=P^\dagger_f \widehat{M_f} P_f$ 
for $f=u$ and $d$, where $\widehat{M_f}$ is given by Eq.~(\ref{M_hat_f}). 
The form of $\widehat{M_f}$ is the most general one based on the $2 \leftrightarrow 3$ symmetry.
In order to reproduce the experimental data, we do fine-tuning the free parameters of the model 
as shown in Fig.~1 and Fig.~2. 
Then, we find that the CKM quark mixing of the model is consistent with the experimental data. 
\par
Let us compare our present model with the mass matrix model~\cite{Koide} 
, which has symmetric $\widehat{M_f}$ and a vanishing (1,1) element  
due to $Z_3$ symmetry and seesaw invariance in addition to 2 \(\leftrightarrow\) 3 symmetry, given by

\begin{equation}
\widehat{M_f}
=
\left(
        \begin{array}{ccc}
         0  & A_{f} &  A_{f} \\ 
        A_{f} & B_{f} & C_{f} \\ 
         A_{f}  & C_{f} & B_{f}
        \end{array}
\right).
\end{equation}
This model, which predicts somehow small value for $|U_{13}|$, is derived from our present model by taking the following $\varphi_{f1}$ and $\varphi_{f2}$ 
which are restricted by the expressions in terms of the quark masses as
\begin{equation}
\varphi_{f2}  = \varphi_{f1}\quad \mbox{and }\quad 
\tan^2(\varphi_{f1}-\alpha )=\frac{-m_{f1}}{m_{f2}}.
\end{equation}
This corresponds to fixing the parameters $\varphi_{d1}$ and $\varphi_{u1}$ as 
$\varphi_{d1}  = 0.395$ and $\varphi_{u1}  = 0.674$, 
if we use the quark mass values\cite{Fusaoka} such that $m_u(M_X)=1.04\, \mbox{MeV}$, 
$m_c(M_X)=302\, \mbox{MeV}$, $m_d(M_X)=1.33\, \mbox{MeV}$, and $m_s(M_X)=26.5\, \mbox{MeV}$. 
Therefore, we find that the values of the $\varphi_{d1}$ and $\varphi_{u1}$ are close to 
the ones we determine by the present model.
Namely, it turns out that the consistent solution of the present model with the data 
is not far from the one of the model with $D_f=0$ and $A^\prime_f=A_f$.





\newpage
\begin{figure}[htbp]
\begin{center}
\includegraphics{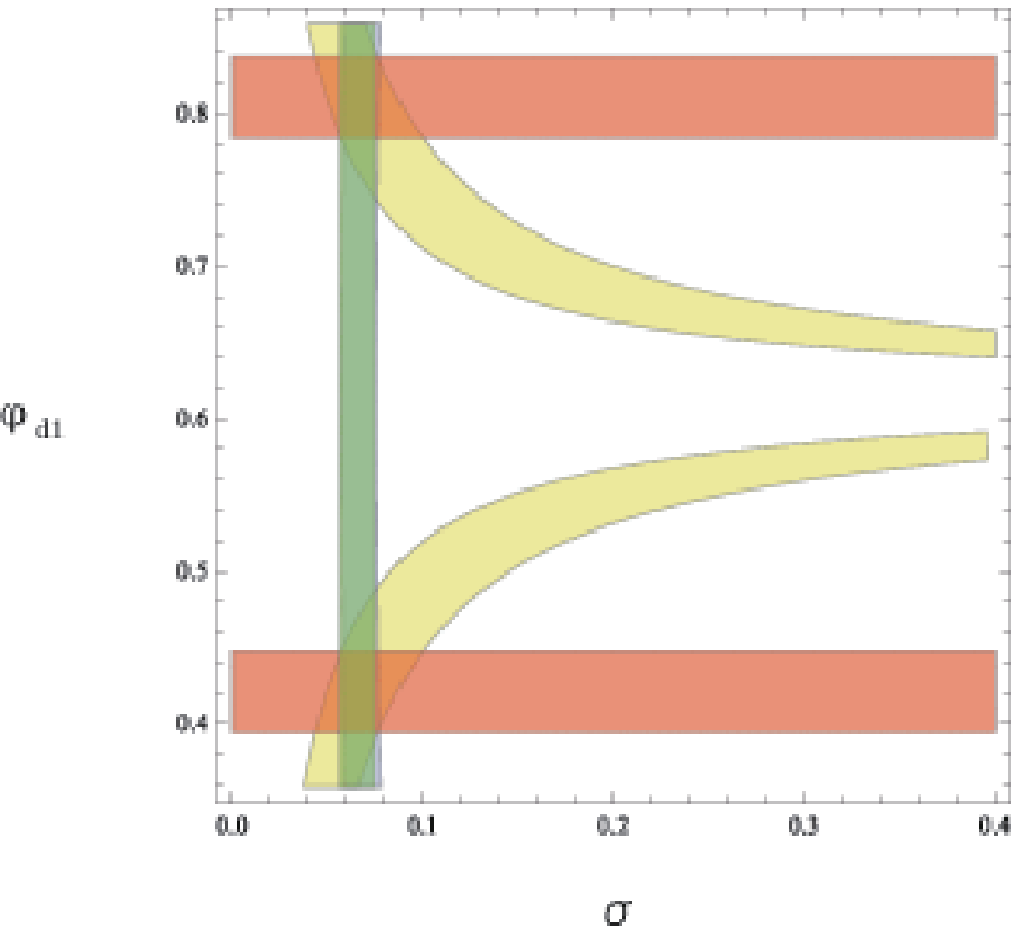}
\end{center}
\caption{%
The allowed region in the $\sigma$ - $\varphi_{d1}$ parameter plane.
Overlapped region is allowed from the experimental data for the CKM quark mixing matrix elements, 
 $|(U_{CKM})_{31}|$(Yellow), $|(U_{CKM})_{32}|$(Blue), 
$|(U_{CKM})_{33}|$(Green), and $|(U_{CKM})_{31}|/|(U_{CKM})_{32}|$(Red).
}
\label{fig1}
\end{figure}

\begin{figure}[htbp]
\begin{center}
\includegraphics[clip,scale=0.68]{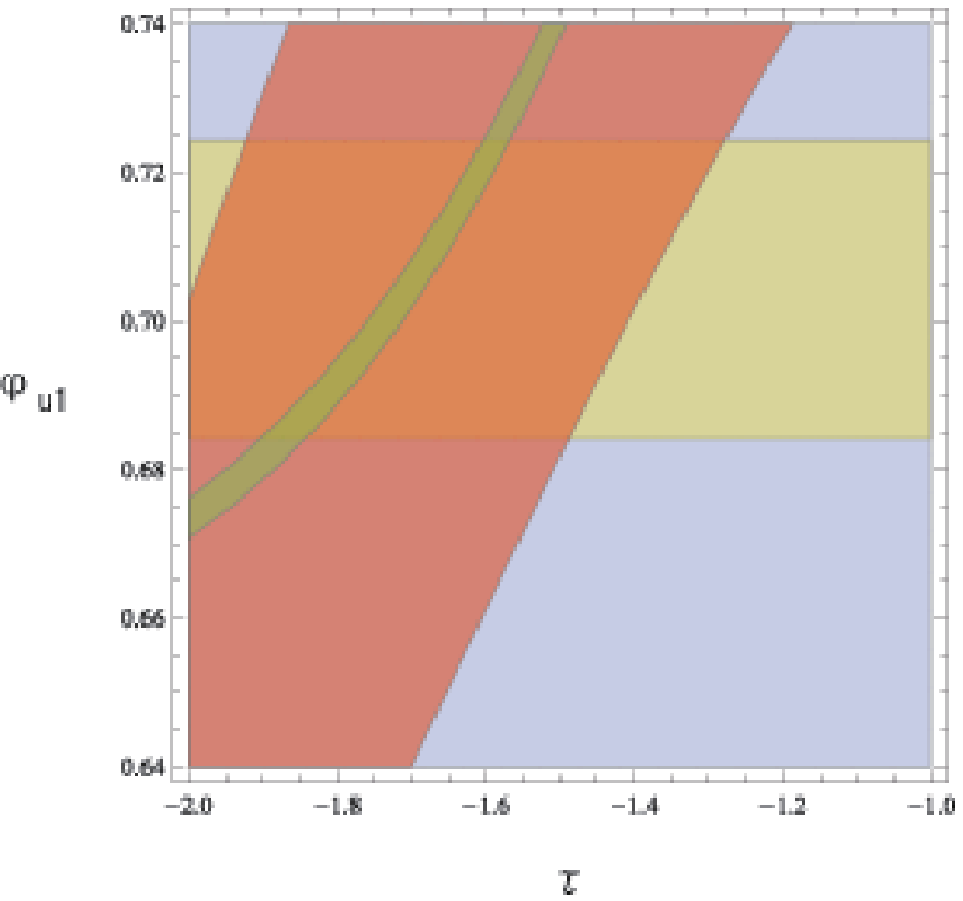}(a)
\includegraphics[clip,scale=0.68]{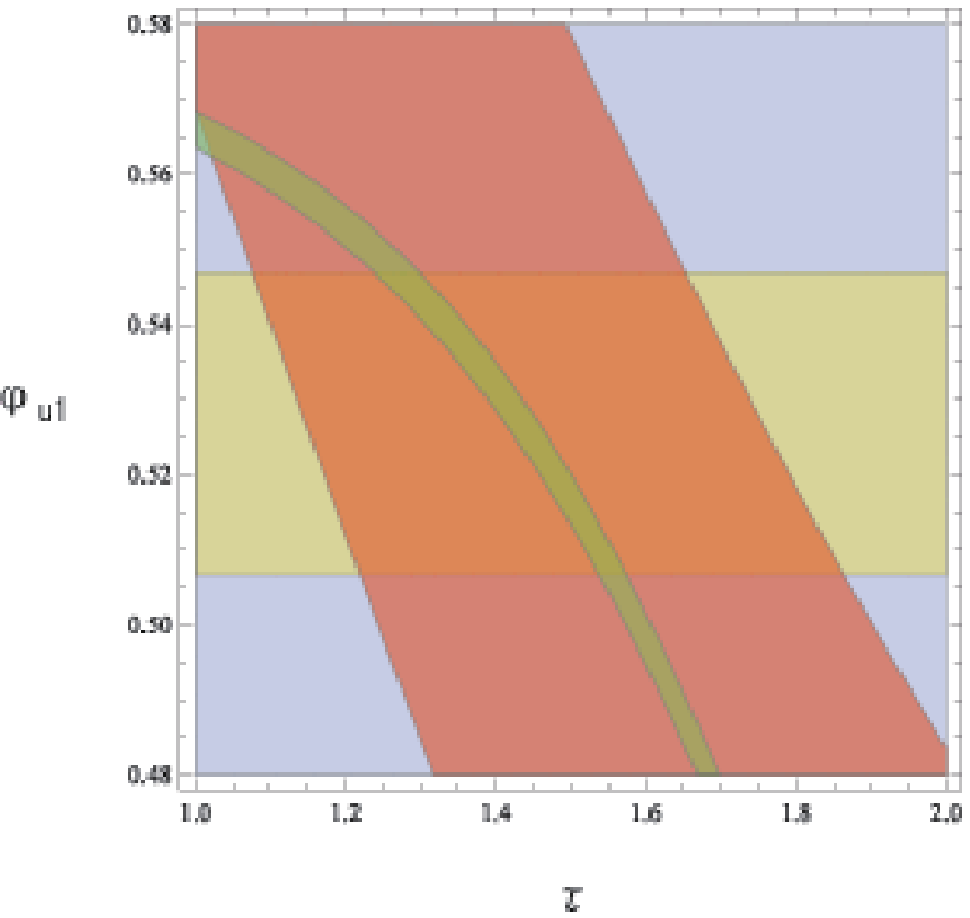}(b)\\ 
\quad\\
\includegraphics[clip,scale=0.68]{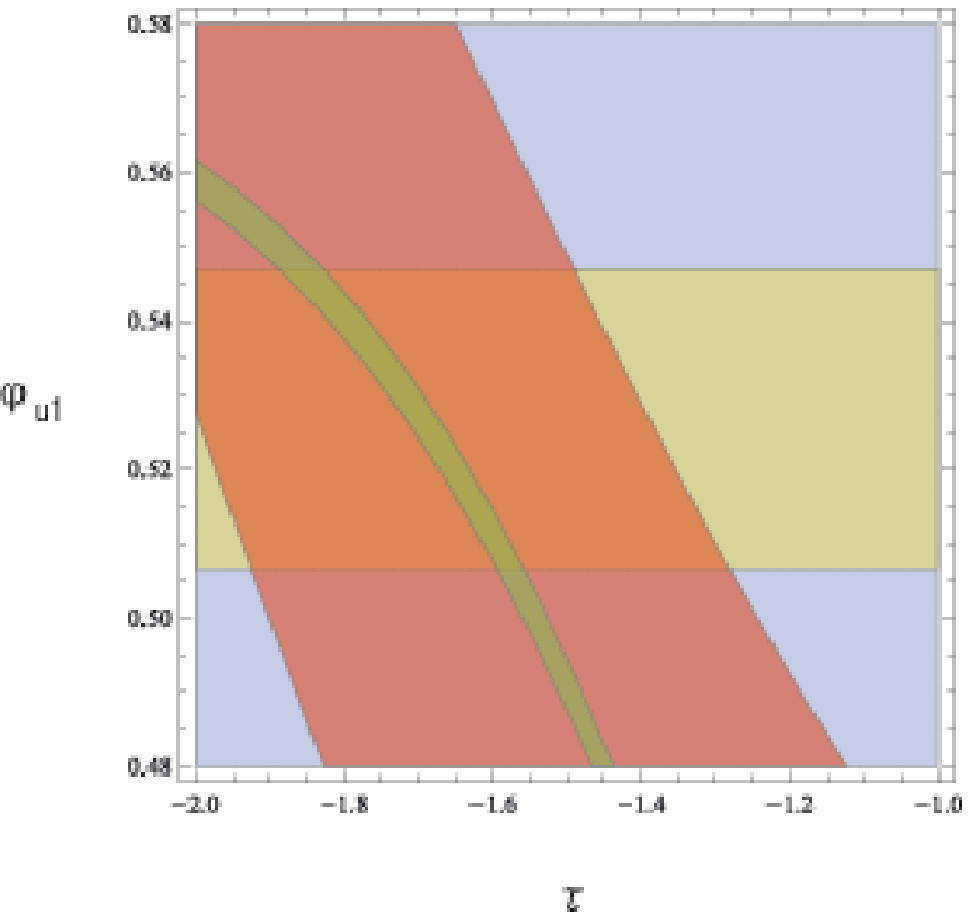}(c)
\includegraphics[clip,scale=0.68]{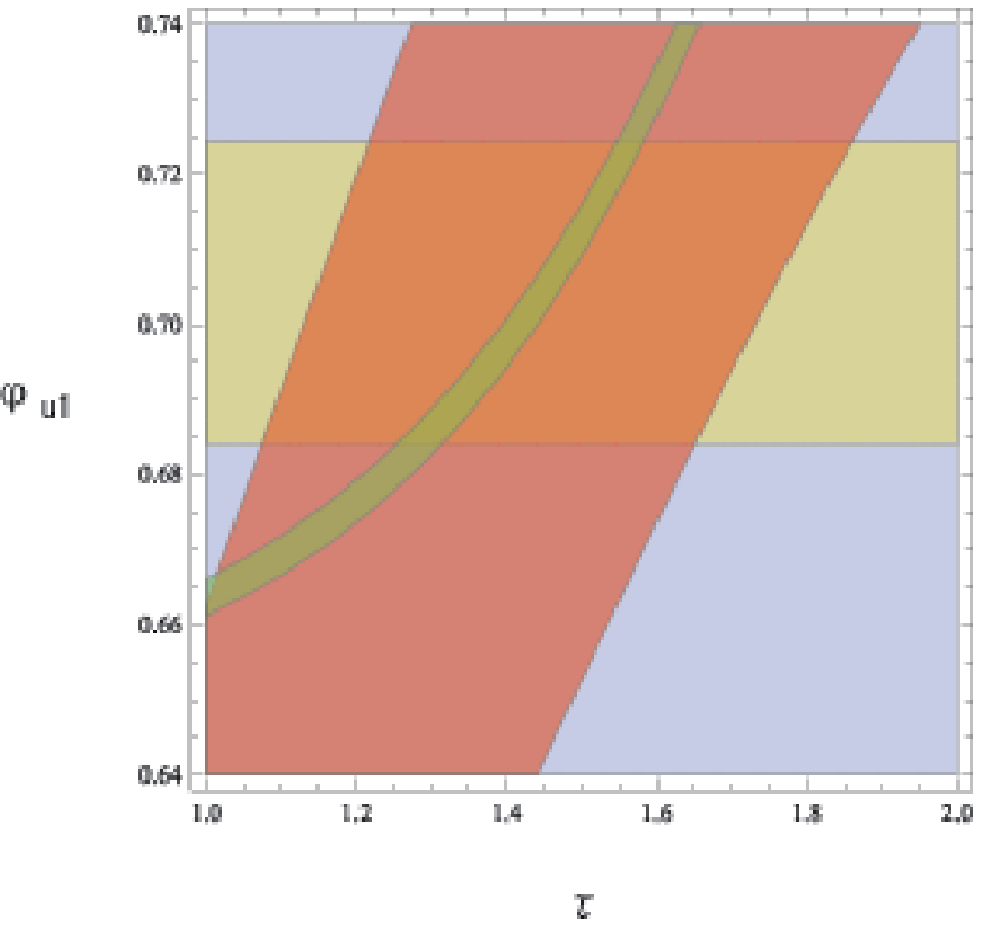}(d)\\
\end{center}
\caption{%
The allowed region in the $\tau$ - $\varphi_{u1}$ parameter plane.
(a) and (b) are shown by taking $\sigma =0.07$ and $\varphi_{d1}=0.42$, while 
(c) and (d) are shown by taking $\sigma =0.07$ and $\varphi_{d1}=0.81$.
Overlapped region is allowed from the experimental data for the CKM quark mixing matrix elements and the $CP$ violating phase, 
 $|(U_{CKM})_{12}|$(Green), 
 $|(U_{CKM})_{13}|$(Yellow), $\delta_q$(Red), and $|(U_{CKM})_{23}|$(Blue). 
}
\label{fig2}
\end{figure}


\begin{thebibliography}{99}
\bibitem{Fukuyama2}
The concept of seesaw invariance was first proposed in \cite{Fukuyama}.
\end{thebibliography}
\end{document}